\documentclass[preprint,12pt]{elsarticle}
\usepackage{graphicx}
\usepackage{subfigure}

\usepackage{amssymb}
\journal{Astroparticle Physics}

\begin{document}

\begin{frontmatter}
\title{Cosmic ray electron anisotropies as a tool to discriminate between exotic and astrophysical sources.}


\author{\bf Ignacio Cernuda}
\ead{ignacio.cernuda@ciemat.es}
\address{Basic Research Division, Centro de Investigaciones Energeticas, Medioambientales y Tecnologicas ({\sc CIEMAT}) Avda. Complutense, 22 - 28040 Madrid,Spain}

\begin{abstract}
Recent results from the PAMELA \cite{Boezio et al. (2008)}, ATIC \cite{ATIC} , PPB BETS \cite{PPBBETS} and Fermi \cite{FermiData} collaborations extend the energy range in the $e^{+},e^{-}$ measurement up to unexplored energies in the hundred GeVs range confirming 
the bump starting at about 10GeV already suggested by HEAT \cite{Heat} and AMS01 data \cite{AMS}. This bump can be explained by annihilating
dark matter in the context of exotic physics, or by nearby astrophysical sources e.g. pulsars. In order to discriminate between competing models for primary positron
production, the study of anisotropies ,complementary to the spectrum determination , shows up as new tool to look for the origin of the lepton excess. In this letter we calculate
the contribution to the electron flux given by the collection of all known gamma ray pulsars (as listed in the ATNF catalogue) and by annihilating dark matter both in case we have a
clumpy halo or in case the excess can be atributed to a nearby sizeable dark matter clump. We address the problem of the electron anisotropy in both scenarios and estimate the prospect 
that a small dipole anisotropy can be found by the Fermi observatory.

\end{abstract}

\begin{keyword}

Dark Matter \sep Anisotropy \sep PAMELA \sep ATIC \sep Fermi/LAT

\end{keyword}

\end{frontmatter}

\section{Introduction}

The recent claim from the PAMELA collaboration of an excess in the ratio of positrons to electrons plus positrons (positron fraction), 
seems to support what HEAT and AMS01  data suggested regarding the existence of a possible primary cosmic ray electrons source. However
some uncertainties remain , e.g the possible enhancement of the positron flux due to the relatively poor knowledge of the propagation parameters set \cite{Delahaye et al. (2008)}.
A large number of candidates have been suggested that may be able to reproduce the observed positron fraction. 
Among the exotic ones, annihilating dark matter may be the most widely studied .
Dark Matter is assumed to annihilate in the galactic halo and its annihilation products should reach the Earth and be measured.
As it has been pointed out in previous works \cite{Cirelli et al. (2008)}, PAMELA data seems to accommodate preferably leptonic channels rather than dark
matter decaying to quark pairs , but typically large boost factors are required to adjust the shape to the data. This boost factors are related to an increase of the annihilation rate due to 
e.g. the recently invoked Sommerfeld effect, annihilation of a non thermal WIMP, or the usual inhomogeneities of the dark matter halo ie, clumps, dark matter mini spikes etc...  Moreover, to accomodate theory to the
existing data, different normalizations are required for the different experiments.
More conventional scenarios can also reproduce the spectral shape.
 Specifically in this work we analyse the possible contribution of nearby astrophysical objects i.e. pulsars 
, and for leptophilic dark matter.\\
In the case of multiple sources, a natural explanation for the spectral shape in
the ATIC data is found, namely the presence of peaks in the spectrum can be reproduced in case we have a convenient selection of pulsars, or in case we have a distribution
of clumps contributing to the total flux with the required luminosity, however it is necessary to fine tune the pulsar parameters to fit the peaks present in the ATIC spectrum.
On the contrary, the Fermi data shows a much flatter spectrum that can be easily reproduced by pulsars with standard parametrizations.
We calculate the predicted dipole anisotropy produced in both scenarios, i.e annihilating dark matter clumps and nearby pulsars, and assess the Fermi detection capability of such
signature at least at the 2 sigma confidence level.

\section{Propagation}

 Cosmic ray propagation is a diffusive process due to the random galactic magnetic fields. If we denote the number of particles of type $i$ per unit volume found in a time t at $\vec{x}$
with energy $E$
by $n_{i}(E,\vec{x},t)$  $[GeV^{-1}cm^{-3}]$, the evolution equation as initially written by $\emph{Ginzburg}$ and $\emph{Syrotvatskii}$ \cite{solut} can be expressed as:

\begin{center}
  \begin{eqnarray}
     \frac{\partial n_i}{\partial t}& = &\nabla (\mathbf{D} \nabla n_i) - \nabla (\vec{v} n_i) - \frac{\partial}{\partial E}(b n_i)  \nonumber \\
     &+ &\beta c N \sum_{s<i} n_s \sigma _{i,s} - \beta c N\sigma _{i}n_i - \frac{n_i}{\gamma\tau _{i}} + Q_i
   \end{eqnarray}
 \end{center}   
 
Where    $\nabla (\mathbf{D} \nabla n_i)$ is the diffusion term with $\mathbf{D} (E,\vec{x} ,t)$ $[{\rm cm}^2/{\rm s}]$ the diffusion tensor;
$\nabla(\vec{v} n_i)$ is the convection term with $\vec{v}=\vec{v}(\vec{x})$ $[{\rm cm}/{\rm s}]$ the velocity of the galactic wind;
$\frac{\partial}{\partial E}(b n_i)$ represents the energy loss with $b(E,\vec{x} ,t)$ $[{\rm GeV}/{\rm s}]$ the rate of energy change;
$\frac{n_i}{\gamma\tau _{i}}$ is the fraction of particles lost by radioctive decay with a characteristic life-time ${\tau}_i$ $[{\rm s}]$;
$\beta c N \sum_{s<i} n_s \sigma _{i,s} - \beta c N\sigma _{i}n_i $ is the number density of particles created or destroyed by spallation processes in the propagation due to
interactions with the interestellar medium of density N, with a
characteristic cross section $\sigma _{i,s}$ $[{\rm cm}^2]$.
Finally we have the source term $Q_i(E,\vec{x} ,t)$ $[{\rm GeV}^{-1}{\rm cm}^{-3}{\rm s}^{-1}]$ that injects particles of type $i$ into the galaxy.
For electron and positron propagation, the relevant processes are diffusion, convection, energy losses and the source term.   
Propagation can be approached in two complementary ways . The first one is to solve the transport equation using numerical methods in the same way the package \emph{GALPROP} 
\cite{Strong et al. (2007)} does.
The second one is to analytically solve the transport equation with a set of realistic simplifying assumptions.
In this work we use the standard \emph{GALPROP} code to get the positron and electron backgrounds and an analytical solution of the propagation equation 
for the primary positron's flux.
For primary positron sources, we have used the Green functions formalism as described in \cite{Green} where the two main processes ,diffusion and energy losses are considered . The resulting diffusion-loss equation for this process is given by:

 \begin{center}
  \begin{eqnarray}
     \frac{\partial}{\partial t}n(E,\vec{x},t)&=&D(E)\cdot \nabla^{2}n(E,\vec{x},t) -  \nonumber \\
     & - & \frac{\partial}{\partial E}(b(E)n(E,\vec {x},t)) + Q(E,\vec{x},t)
   \end{eqnarray}
   \label{diff-eq}
 \end{center}
 
Where  $b(E)=-\frac{dE}{dt}= aE^{2} + bE + c \approx aE^{2}$ codifies the energy losses due to (a) inverse Compton and synchrotron radiation , (b) bremsstrahlung  and (c) ionisation.
At the energies we are interested $\rm{E}> 10$ GeV the energy loss is very well aproximated by the synchrotron losses in the interstellar (ISM) magnetic fields and inverse Compton off the CMB
and starlight at optical and IR frequencies.
The energy loss due to these processes is calculated as \cite{solut}:

\begin{center}
 \begin{eqnarray}
 -\frac{dE}{dt}&=&  \frac{32\pi}{9}c\left(\frac{e^{2}}{mc}\right)^{2}\left(w_{0}+ \frac{B^{2}}{8\pi}\right)\left(\frac{E}{mc^{2}}\right)^{2} \nonumber \\
               &=& 8\rm{x}10^{-17}\left(w_{0} +6\rm{x}10^{11}\frac{(B/1G)^{2}}{8\pi}\right)E^{2} \approx 10^{-16}E^{2}[{\rm{GeV} \rm{s}}^{-1}]
  \end{eqnarray}
 \end{center}   
 
where the energy density $w_{0}=w_{CMB}+w_{opt-IR}$ is the energy density of the photon background [$eV/cm^{3}$] with $w_{CMB}=0.25 eV/cm^{3} \;\rm{and}\; w_{opt-IR}=0.5 eV/cm^{3}$ . $w_{B}$ stands for the ISM magnetic field energy density
$w_{B}=\frac{B^{2}}{8\pi}=0.6 eV/cm^{3}$ for B=$5\mu$G.\\
For the diffusion coefficient $D(E)=D_{0}E^{\delta}$, three setups MAX,MED and MIN can be considered \cite{Delahaye et al. (2008)} which are consistent with the B/C .

\begin{table}[ht]
\caption{Diffusion Setups}
\centering 
\begin{tabular}{c c c } 
\hline\hline %inserts double horizontal lines
 & $D_{0}$  [$cm^{2}/s$] & $\delta$  \\ [0.5ex] % inserts table
\hline 
MAX & $1.8\;\rm{x}10^{27}$ & 0.55  \\ 
MED & $3.4\;\rm{x}10^{27}$ & 0.70  \\
MIN & $2.3\;\rm{x}10^{28}$ & 0.46   \\ [1ex]
\hline 
\end{tabular}
\end{table}

We solve the equation for a steady state source (DM positron injection) and for a nonstationary source (SNRs and pulsars) assuming free boundary conditions.

\section{Astrophysical sources of high energy positrons}

Among the astrophysical objects that populate our Galaxy, many can contribute to the positron abundance in cosmic rays, but the required energy excludes a large part of them.
Following a Hillas argument, the astrophysical sources able to inject the required order of energy can be found in e.g: SNRs and pulsars .
Gamma Ray pulsars are expected to produce pairs of electrons and positrons as a result of electromagnetic cascades induced by acceleration of electrons in the magnetosphere.
The accelerated electrons emit curvature radiation i.e. photons, above the pair creation threshold by magnetic conversion. Positron injection can be expressed as a power law
with an exponential cutoff at $E_{c}$ , $\frac{dN_{e}}{dE}=E^{-\alpha}e^{-E/Ec}$. Previous works (e.g. \cite{Busching et al. (2008)},\cite{Hooper-Blasi-Serpico (2008)},\cite{Yuksel et al.
(2008)},\cite{Harding et al.},\cite{Chi et al},\cite{Gao et al. (2008)},\cite{Profumo (2008)},\cite{Grasso et al. (2009)} ) have shown the plausibility of the pulsar scenario as sources of primary cosmic ray electrons.
In our analysis, we will assume a benchmark model as considered in \cite{Profumo (2008)}, namely, the "standard model" (ST). Although more refined models for electron production in
pulsars can be considered, some of them (e.g. Harding-Ramaty model \cite{Harding et al.} or the one devised by Zhang and Cheng \cite{Zhang et al. (2001)} ) produce an $e^{\pm}$
output well below the
observations or a comparable one (e.g. Chi et al. \cite{Chi et al}). In the latter case, the most outstanding pulsars produce similar patterns to those considered in the ST model, so we
will assume the most simple scenario for positron production in pulsars as a benchmark model for the study.\\
The ST model assumes that all the rotational energy of the pulsar is lost through magnetic dipole radiation. 
As the rotational energy is given by $E=I\Omega^{2}/2$ (where I$\approx 10^{45}g cm^{2}$ is the moment of inertia and $\Omega$ the spin frequency), the spindown power will
be $\dot{E}=I\Omega\dot{\Omega}$. 
For such a magnetic dipole radiator, the energy loss rate can be written as a function of the neutron star radius R, and $\alpha$ the angle between the dipole axis and rotation $\dot{E}=-\frac{B^{2}R^{6}\Omega^{4}\sin^{2}{\alpha}}{6c^{3}}$ , i.e. $\dot{\Omega}\varpropto -\Omega^{3}$. 
Integrating this expression leads to the solution of the rotational velocity of a pulsar where the magnetic dipole radiation braking dominates: 

 \begin{equation}
  \Omega(t)=\frac{\Omega_{0}}{(1+t/\tau_{0})^{1/2}}
 \end{equation}

where $\tau_{0}=\frac{3c^{3}I}{B^{2}R^{6}\Omega_{0}^{2}}$ is a characteristic time taken to be around $10^{4}$ years for nominal pulsar parameters.
The luminosity of the pulsar can then be derived as:

\begin{equation}
  L(t)=\frac{L_{0}}{(1+t/\tau_{0})^{2}}
 \end{equation}
 \label{lumin}

Integrating this expresion over the pulsar age T, the total energy output can be approximated by $E_{out}=I\int dt \Omega\dot{\Omega}\approx I\Omega_{0}/2$. If $t/\tau_{0}>>1$ then
$\Omega_{0}^{2}\simeq \Omega^{2}\frac{t}{\tau_{0}}$ resulting in an energy output into electrons of:

 \begin{equation}
  E_{out}[ST]=f_{e^{\pm}}\dot{E}\frac{T^{2}}{\tau_{0}}
 \end{equation}
 
 where the energy budget is determined by the spin down power \.{E}, the age of the source T and the conversion efficiency  into pairs  $f_{e^{\pm}}$ that is assumed to be of a
 few \%. 
 Thus, the $e^{\pm}$ source for a single pulsar located at a distance r, injecting positrons at time t and energy E can be expressed as:
 \begin{equation}
 Q(E,r,t)=N\cdot L(t)E^{-\alpha}\exp{(-E/E_c)}\delta(r)
 \end{equation}
 
 where N is the normalization factor taken to satisfy the total energy release constraint 
 $E_{tot}=\int^{T}d\tau\int^{e_{max}}_{1GeV} E Q(E,\tau)dE$ and L is the luminosity of the source. We have introduced an spectral cutoff at $E_{c}$=1TeV motivated by the ATIC and
 Hess data. This cutoff will be relevant for young pulsars, as old pulsars have a maximal $e^{\pm}$ energy below the cutoff due to energy losses.
 In this work we have assumed typical pulsar injection $L(\tau)=\frac{L_{0}}{(1+\frac{\tau}{\tau_{0}})^{\frac{n+1}{n-1}}}$  with the usual braking index n=3 for magnetic dipole
 radiation braking, resulting in Eq. \ref{lumin}, nonetheless a similar analysis can be conducted 
 for other choices i.e. exponential decay luminosity as it is expected from microquasars or $e^{\pm}$ release from the nebula that surrounds pulsars.
 
 For this purpose we will consider all the gamma ray pulsars listed in the Australian Telescope National Facility (ATNF) \cite{Manchester et al. (2004)} pulsar catalogue
 \footnote{http://www.atnf.csiro.au/research/pulsar/psrcat}.
 Young pulsars with typical ages lower than $10^{4}-10^{5}$ years are considered to be surrounded by the pulsar wind nebula (PWN) or a SNR shell that confines the injected
 electrons before releasing them to the ISM. This has to be taken into account when we consider the age of the pulsars that can contribute to the electron abundance. In this 
 respect we will consider two collections. In the first one we will take a lower bound of $10^{4}$ years constraining our pulsar collection to ages between $10^{4}$ and $10^{7}$ years. This constraint allows us to accept pulsars like Vela which
 most probably are still surrounded by the PWN. In this scenario, we assume a low conversion efficiency for the young pulsars ($\mathcal{O}(1\%)$) to take into account the posible confinement of
 leptons. In the second one, we will consider that pulsars with ages lower than $5\rm{x}10^{4}$ years cannot contribute to the bulk of
 electrons, i.e. we constrain our collection to mature pulsars. This introduces an injection delay $\Delta t$ between the pulsar birth and injection into the ISM due to the confinement of the electrons in the PWN. This
 delay may be important for young pulsars for wich $T\simeq \Delta t$ but for sufficiently old pulsars we can safely dismiss the delay issue and set the injection time at the
 pulsar age. \\
 The condition for gamma ray emmision is taken to be as \cite{Zhang et al. (1997)} the fraction size of the outer gap $g=5.5 P^{26/21}B_{12}^{-4/7} < 1$ in terms of the pulsar period P and
 the pulsar surface magnetic field $B_{12}$ (in $10^{12}G$ units ) resulting in a collection of more than one hundred pulsars, from
 which 3 lie at a distance $<$ 1kpc.
 In the determination of the injection spectral index $\alpha$, we have to take into account the constraints that come from 
  observations of synchrotron radiation from SNRs. We could also assume, as in a Harding-Ramaty model \cite{Profumo (2008)}, that the $e^{\pm}$ have the same spectral index as
  gamma rays emmited by pulsars 
  (which has been measured by EGRET to be around 1.4-2.2 for energies $0.1 < E < 10 GeV$). Altogether, we shall consider spectral indexes from 1.4 to 2.2 . 
 For the sake of simplicity we will assume that our collection of pulsars have all the same spectral index $\alpha =1.7$ and 
 as an additional simplifying assumption we will take a universal set of parameters for the whole collection. In this way, neither PAMELA nor ATIC can be reproduced properly, but with a little
 of fine tuning work, the spectral features can be achieved. For example, as recently pointed out by \cite{Kawanaka et al. (2009)} , the width of the peak produced by continuous
 injection depends on the characteristic time of the luminosity of the pulsar by $\frac{\Delta\epsilon_{peak}}{\epsilon_{peak}}\simeq\frac{\tau_{0}}{T}$. Proceeding like this, selecting a number of well known pulsars and adjusting their luminosity parameters,
 the spiky spectral shape of ATIC can be achieved without violating the PAMELA constraints. Additionally, the energy losses set up the age of the
 source that produces the peak around 600 GeV reported by ATIC, provided that it is far ($\sim 1Kpc$) and bright enough ($\sim$ 1 order of magnitude in the conversion efficiency). 
 As already noted by \cite{Profumo (2008)}, PSR B0355+54 fulfills the requirements. For this pulsar, a very large conversion efficiency into pairs is required to account for the ATIC
 peak ($\mathcal{O}(40\%)$). 
 On the other hand, the much more statistically significant Fermi/LAT data shows a much flatter spectrum. Just taking standard pulsar parameters is enough to fit the data without
 having to resort to very large conversion efficiencies so this is the approach we will follow, although it should be noted that due to the poor Fermi energy resolution at these energies , actually existing peaks could be smoothed
 down to the observed spectrum, proving the issue of normalization of every pulsar to be a subtle one .\\
 Once we have the positron source we proceed to calculate the number density of positrons by solving the diffusion-loss equation (Eq.\ref{diff-eq}) for a non stationary source . The solution of the
 equation has been previously derived for a burstlike power law injection source with a cutoff Ec (\cite{Atoyan et al. (1995)}, \cite{Grasso et al. (2009)}) .
 
  \begin{center}
   \begin{equation}
   \phi(E,r,t)=\frac{\beta c}{4\pi}\frac{Q_{0}}{\pi^{3/2}r^3}\left( \frac{r}{D_{diff}} \right)^{3} (1-atE)^{\alpha -2}E^{-\alpha}e^{- \frac{E}{(1-atE)E_{c}} } e^{-\left(\frac{r}{D_{diff}}\right)^2}
   \end{equation}
 \end{center}

where the distance scale is aproximately,

   \begin{center}
   \begin{equation}
   D_{diff}(E,t)\simeq 2 \sqrt{D(E) t \frac{1-(1-E/E_{max})^{1-\delta}}{(1-\delta)E/E_{max}} }
   \end{equation}
 \end{center}
 
  as a function of the diffusion index $\delta$ and the maximum energy given by the energy losses: $E_{max}\simeq 1/at$  with $a\simeq 10^{-16} \; GeV^{-1}s^{-1}$ .\\ 
 
\begin{figure}[!hb]
 \begin{center} 
  \includegraphics[width=0.7\columnwidth]{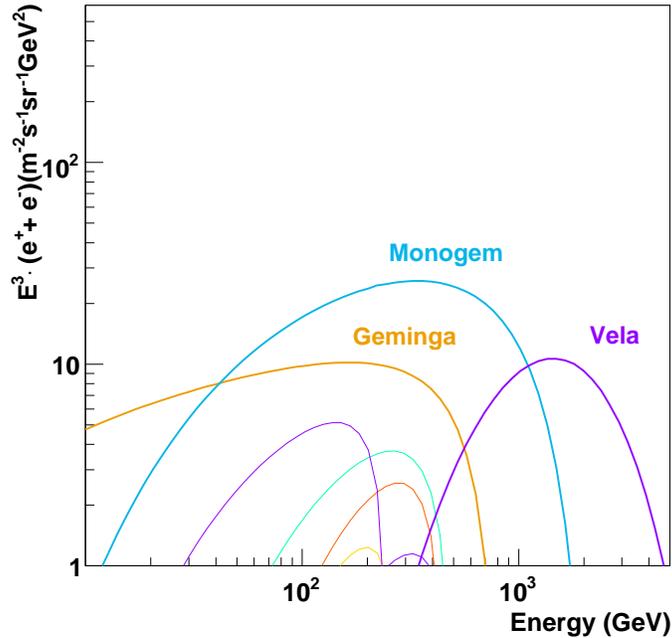}  
 \caption{Contribution to the electron flux injected by the collection of pulsars considered in the BM2 setup.}
 \label{pulsars_ind}
  \end{center}
  \end{figure}

  As the source emits with luminosity $L(t)$, the flux will be given by $\phi(E,r,t)=\int_{T}^{t}L(t')\phi(E,r,t')dt'$.
  We have calculated the pulsar contribution to the local electron flux in the case of burstlike injection and continuous injection for our collection of young and mature pulsars.
  Even in the continuous case , due to the steep of the index of dipolar emission, the injection is well aproximated by a burstlike event, being the effect of
  following an approach of continuous injection to soften the posible contributions of the individual pulsars. 
  In order to reproduce the spectral features of the Fermi data, we assume a MED diffusion scenario with an overall conversion efficiency of 3\%. Due to the relative variability
  of the collection of pulsars in the age/distance parameter space, we adjust the conversion efficiency for a few number of objects that show prominent features in the spectrum at the
  considered energies, namely, Geminga, Monogem and J2043+2740 ( Fig. \ref{pulsars_ind} ).

   We can consider two sets of pulsars (BM1,BM2) to address the lepton confinement in the PWN. Our first set will be made up of mature pulsars
  with ages T$>5\rm{x}10^{4}$ years, thus , introducing a delay between the pulsar birth and the electron release to the interstellar medium of $\Delta t\simeq 5\rm{x}10^{4}$years.
  In our second set, we consider also younger pulsars, including Vela, with ages T$>10^{4}$ years and a delay of $\Delta t\simeq 10^{4}$ years. In this scenario, as young pulsars are still
  surrounded by the nebula that confines the electrons, a smaller efficiency must be called for. The efficiency we will assume for this pulsars, including Vela, is
  $\mathcal{O}(1\%)$, and is taken to satisfy the bounds imposed by the Hess data. A larger efficiency would imply a non neglible contribution above the background of secondary
  electrons around 2 TeV due to the contribution of Vela, which has not been not observed.

%\begin{table}[ht]
%\caption{Pulsar Parameters}
%\centering 
%\begin{tabular}{|c|c|c|c|c|} 
%\hline %inserts double horizontal lines
% Pulsar & d[Kpc] & T [$10^{5}$ years]  &  $E_{out}$[ST] BM1/BM2 [$10^{50}$GeV]& $f_{e^{\pm}}$  BM1/BM2 [\%] \\ [0.5ex] % inserts table
%\hline 
%Geminga & 0.16 & 3.42 & 7.4/4.4  & 10.0/6.0 \\ \hline 
%Monogem & 0.29 & 1.11 & 1.7/2.83  & 18.0/30.0  \\\hline 
%J2043+2740 & 1.13 & 12.0 & 1.6/1.6  &  0.1/0.1  \\ \hline 
%Vela & 0.29 & 0.11 & 0.0/0.17  &  0.0/1.0  \\ %[1ex]
%\hline 
%\end{tabular}
%\end{table}

\begin{table}[h]
\caption{Pulsar Parameters}
\centering 
\begin{tabular}{|c|c|c|c|c|} 
\hline %inserts double horizontal lines
 Pulsar & d[Kpc] & T [$10^{5}$ years]  &  $E_{out}$ BM1/BM2 [$10^{50}$GeV]& $f_{e^{\pm}}$  BM1/BM2 [\%] \\ [0.5ex] % inserts table
\hline 
Geminga & 0.16 & 3.42 & 7.4/4.4  & 10.0/6.0 \\ \hline 
Monogem & 0.29 & 1.11 & 1.7/2.83  & 18.0/30.0  \\\hline 
J2043+2740 & 1.13 & 12.0 & 1.6/1.6  &  0.1/0.1  \\ \hline 
Vela & 0.29 & 0.11 & 0.0/0.17  &  0.0/1.0  \\ %[1ex]
\hline 
\end{tabular}
\end{table}

 \begin{figure}[!ht]
 \centering
   \subfigure[]{\includegraphics[width=0.47\textwidth]{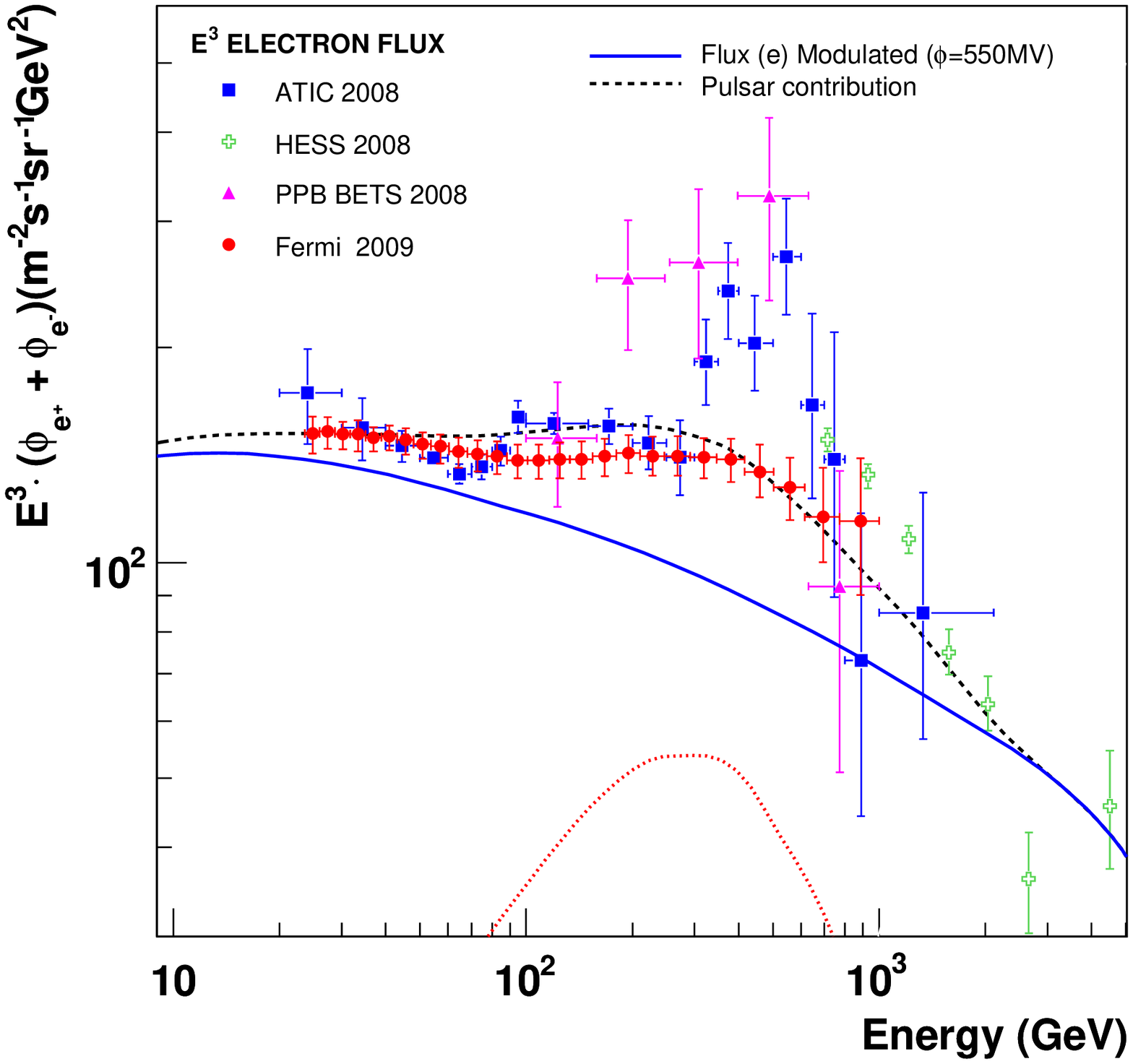}}
   \subfigure[]{\includegraphics[width=0.47\textwidth]{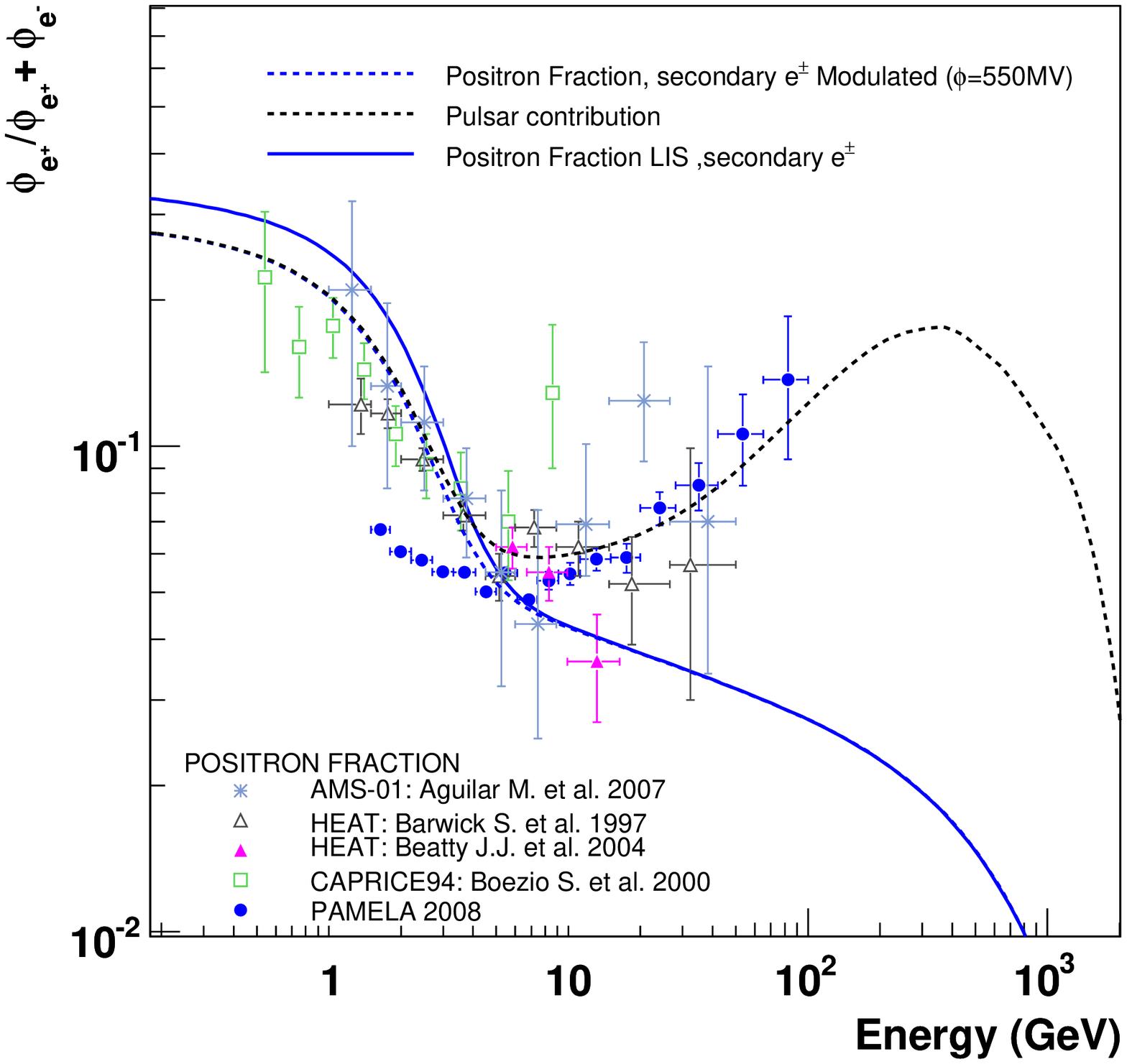}}   
   \caption{Mature pulsar contribution to the electron flux for a MED diffusion setup and overall efficiency $f_{e^{\pm}}=3\%$ . We assume an injection index of $\alpha=1.7$ and a
 cutoff $E_{c}=1TeV$. We consider burstlike injection and the efficiencies have been adjusted to $f_{\pm e}^{Monogem}=18\%$, $f_{\pm e}^{Geminga}=10\%$, $f_{\pm e}^{J2043+2740}=0.1\%$ to reproduce the Fermi and PAMELA data. 
 (a): $E^{3}\cdot (e^{+}+e^{-})$ (b): Positron Fraction }
   \label{pulsars} 
  \end{figure}
  
    \begin{figure}[!hb]
 \centering
   \subfigure[]{\includegraphics[width=0.47\textwidth]{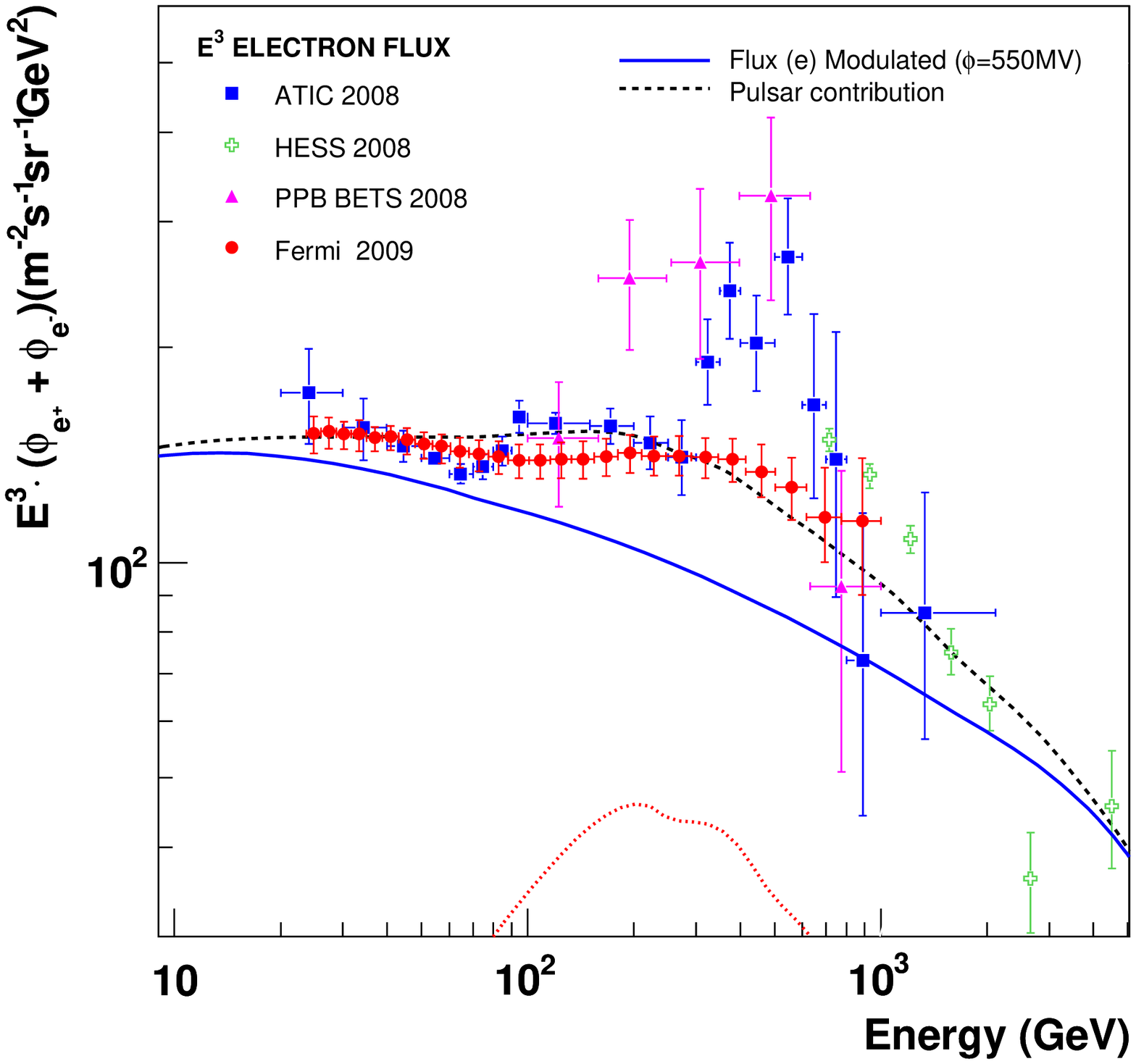}}
   \subfigure[]{\includegraphics[width=0.47\textwidth]{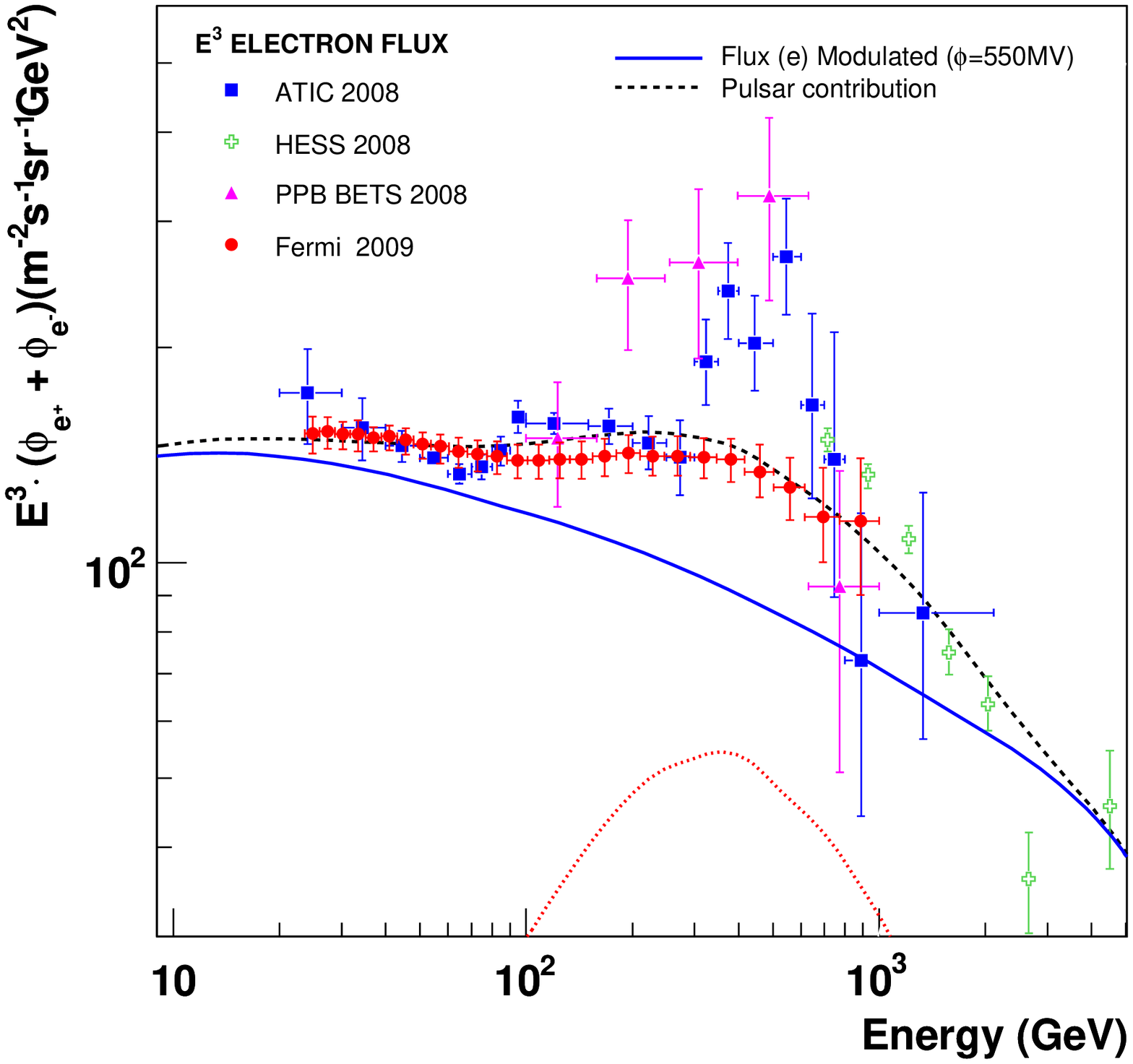}}  
 \caption{(a):Mature + young pulsar contribution to the $e^{+}+e^{-}$ flux for a MED diffusion setup , burstlike injection and $f_{e^{\pm}}=3\%$ . We assume an injection index of $\alpha=1.7$ and
 a cutoff $E_{c}=1TeV$.The efficiencies have been adjusted to $f_{Monogem}^{\pm e}=30\%$, $f_{Geminga}^{\pm e}=6\%$, $f_{J2043+2740}^{\pm e}=0.1\%$, $f_{Vela}^{\pm e}=1\%$ to reproduce the Fermi data.
 (b): Mature pulsar contribution to the $e^{+}+e^{-}$ flux for a MED diffusion setup , continuous injection and overall efficiency $f_{e^{\pm}}=3\%$ . We assume an injection index of $\alpha=1.7$ and
  a cutoff $E_{c}=1TeV$. The efficiencies have been adjusted to $f_{Monogem}^{\pm e}=15\%$, $f_{Geminga}^{\pm e}=6\%$, $f_{J2043+2740}^{\pm e}=0.1\%$ to reproduce the Fermi data.}
  \label{young}	
  \end{figure}
  
The resulting fluxes are shown ( Figs. \ref{pulsars} and \ref{young} ) for the considered scenarios. In order to account for the experimental data in the hundreds of GeV, the conversion efficiency into
pairs of Monogem and Geminga must be above the nominal value of 3\%  but inside the standard range (1\%-30\%) considered in \cite{D.Malyshev et al (2009)}. This values may change
depending on the diffusion setup considered, the PWN $e^{\pm}$ release delay, pulsar cutoff etc... so it should be considered as a single realisation of the multiple possibilities that
can reproduce the data \cite{Grasso et al. (2009)}. 

\section{Dark Matter source of high energy positrons}

 According to the standard $\Lambda$CDM cosmological model, a $\simeq 22\% $ of the energy content of the universe is in the form of cold dark matter (CDM). Probably the leading
 candidates to account for it are weakly interacting particles (WIMPS), with the neutralino and the kaluza-klein boson $B^{1}$ the most extensively studied
 ones. The relic density of these particles is determined by their annihilation cross section. Observations of the CMB and large scale structure surveys estimate the relic density 
 constraining the annihilation cross section up to a canonical value of $\langle \sigma v\rangle \approx 3\cdot 10^{-26}cm^{3}s^{-1}$ .
 This value sets the rate of production of standard model particles , e.g leptons, that can be measured as tracers of dark matter annihilation.\\ 
 The signal that results from DM annihilation, depends on the squared DM density
 from the astrophysical side, and on the DM particle mass and cross section from the particle physics side. Since the annihilation rate depends on the squared density, the
 presence of clumpiness or substructure imply an enhancement of the signal compared to a smooth density distribution.
 The present dark matter structure is considered to have its roots in small amplitude quantum fluctuations during inflation. In the accepted "bottom-up" hierarchical structure
 formation, smaller clumps gather together to form larger systems , completely determined by the initial power spectrum of the primordial fluctuations. 
 Galaxies are thus embedded in large dark matter halos that in turn are made up of self-bound substructure or subhalos. The mass distribution, abundance and internal structure of
 clumps is determined by means of high-resolution numerical simulations as the one conducted by Diemand et al.\cite{Diemand et al. (2006)}.
 In this work we make use of the mass distribution of clumps that results from Diemand's simulation as in Cumberbatch et al. \cite{Cumberbatch et al. (2006)} expressed as:
 
  \begin{equation}
  \frac{dn_{D}}{d\log(M/M_{\odot})}\;\;\alpha \;\;(M/M_{\odot})^{-1}\exp[-(M/M_{cutoff})^{-2/3}]
  \end{equation}

Where the lower mass cutoff is $M_{cutoff}\simeq 8.03 \rm{x} 10^{-6}M_{\odot}$ and, inspired by Diemand et al. simulation , we take an upper mass cutoff at $10^{10}M_{\odot}$.
We normalise the distribution such that we have a local clump density (r=$r_{\odot}=8.5 Kpc$) of 500 $pc^{-3}$ between $10^{-6}M_{\odot}$ and $10^{-5}M_{\odot}$. 
We assume for the spatially dependent number density a spherically symmetrical Navarro,Frenk\&White (NFW) profile so the density of clumps is given by

  \begin{equation}
  \frac{dn(r,M)}{d\log(M/M_{\odot})} = \frac{\rho_{0}}{(r/R)[1+(r/R)]^{2}} \frac{dn_{D}}{d\log(M/M_{\odot})}
  \end{equation}
  
Where R=20Kpc and $\rho_{0}=0.86$ for a correct normalization. 
  
For the internal structure within a clump, we adopt a NFW density profile that gives a reasonable fit to the lightest clumps in Diemand's simulation and reproduces well the outer
parts of the large scale subhalos.

\begin{equation}
  \rho(r)=\frac{\rho_{0}}{c(r/r_{200})[1+c(r/r_{200})]^{2}}
 \end{equation}
 
where the maximum radius of the clump is defined as the radius $r_{200}$ where the density equals 200 times the critical density. 
We assume a universal concentration parameter c for all the clumps, found to lie in Diemands simulation within the $1.6 \leq c \leq 3.0$ range, and a constant density core below
$10^{-9}$Kpc.

With all these considerations, the source term can be expressed as:

 \begin{eqnarray}
  Q(E,r)=\frac{\langle\sigma v\rangle}{m^{2}_{DM}}&&f(E)\int^{M_{max}}_{M_{min}}f^{2}_{NFW}(c,M)\nonumber \\
  &&\frac{dn(r,M)}{d\log(M/M_{\odot})}d\log(M/M_{\odot})
  \end{eqnarray}

where f(E) denotes the number of positrons generated per annihilation and energy interval (for each branching channel) and $f^{2}_{NFW}(c,M)$ is the integrated squared density for each DM clump

\begin{equation}
 f^{2}_{NFW}(c,M)=\int^{r_{200}(M)}_{0}4\pi r^{'2}\rho^{2}(r^{'})dr^{'}
  \end{equation}

Once we have the source term we proceed to calculate the flux at the Earth by solving the steady state ($\frac{\partial}{\partial t}n(E,\vec{x})=0$) of Eq.\ref{diff-eq} as already
done e.g in \cite{Green} or \cite{Edsjo}. 
The Eq.\ref{diff-eq} can be greatly simplified translating the energy into s(E).

 \begin{center}
   \begin{eqnarray}    
    s(E)\, & = & \, \int_{E}^{\infty}\frac{\rm{D(x)}}{\rm{b(x)}}\rm{dx}
   \end{eqnarray}
 \end{center}

The equation then reads:

  \begin{center}
  \begin{equation}
     \Delta \psi (s,\vec{x}) -  \frac{\partial}{\partial s}\psi(s,\vec {x})=  - \bar{Q}(s,\vec{x})
   \end{equation}  
 \end{center}

where $\psi(E,\vec {x}) = b(E)\cdot n(E,\vec {x})$ and $\bar{Q}(E,\vec{x})=\frac{b(E)}{D(E)}Q(E,\vec{x})$.

The solution , is given by the convolution of the Green function over the sources:

   \begin{center}
   \begin{equation}
    \phi(E,\vec{x})= \frac{\beta c}{4\pi}\int G(E,\vec{x},s_0,\vec{x}_0)Q(s_0,\vec{x}_0)ds_0d\vec{x}_o
   \end{equation}
 \end{center}

where the Green function is given by :

 \begin{center}
   \begin{equation}
   G(E,\vec{x},s_0,\vec{x}_0)=\frac{b[E(s_0)]}{D[E(s_0)]}\frac{e^{-(\vec{x}-\vec{x_0})^2/D_{diff}^{2}}}{[\pi D_{diff}^2]^{3/2}}
   \end{equation}
 \end{center}

as a function of the integrating variable $s_0$. $D_{diff}$ is the diffusion distance scale given by  $D_{diff}=(4(s(E)-s(E_0)))^{1/2}$ in terms of the variable s(E) with E the
observed energy and $E_{0}$ the energy at injection.

%The solution , is given by the convolution of the Green function over the sources:

%   \begin{center}
%   \begin{equation}
%    \phi(E,\vec{x})= \frac{\beta c}{4\pi}\int G(E,\vec{x},s_0,\vec{x}_0)Q(s_0,\vec{x}_0)ds_0d\vec{x}_o
%   \end{equation}
% \end{center}
  
%where the Green function is given by :

% \begin{center}
%   \begin{equation}
%   G(E,\vec{x},s_0,\vec{x}_0)=\frac{b[E(s_0)]}{D[E(s_0)]}\frac{e^{-(\vec{x}-\vec{x_0})^2/D_{diff}^{2}}}{[\pi D_{diff}^2]^{3/2}}
%   \end{equation}
% \end{center}

%and $D_{diff}$ is the diffusion distance scale given by  $D_{diff}=(4(s(E)-s(E_0)))^{1/2}$ in terms of the variable s(E)

% \begin{center}
%   \begin{eqnarray}    
%    s(E)\, & = & \, \int_{E}^{\infty}\frac{\rm{D(x)}}{\rm{b(x)}}\rm{dx}
%   \end{eqnarray}
% \end{center}

\begin{figure}[!hb]
 \centering
   \subfigure[]{\includegraphics[width=0.47\textwidth]{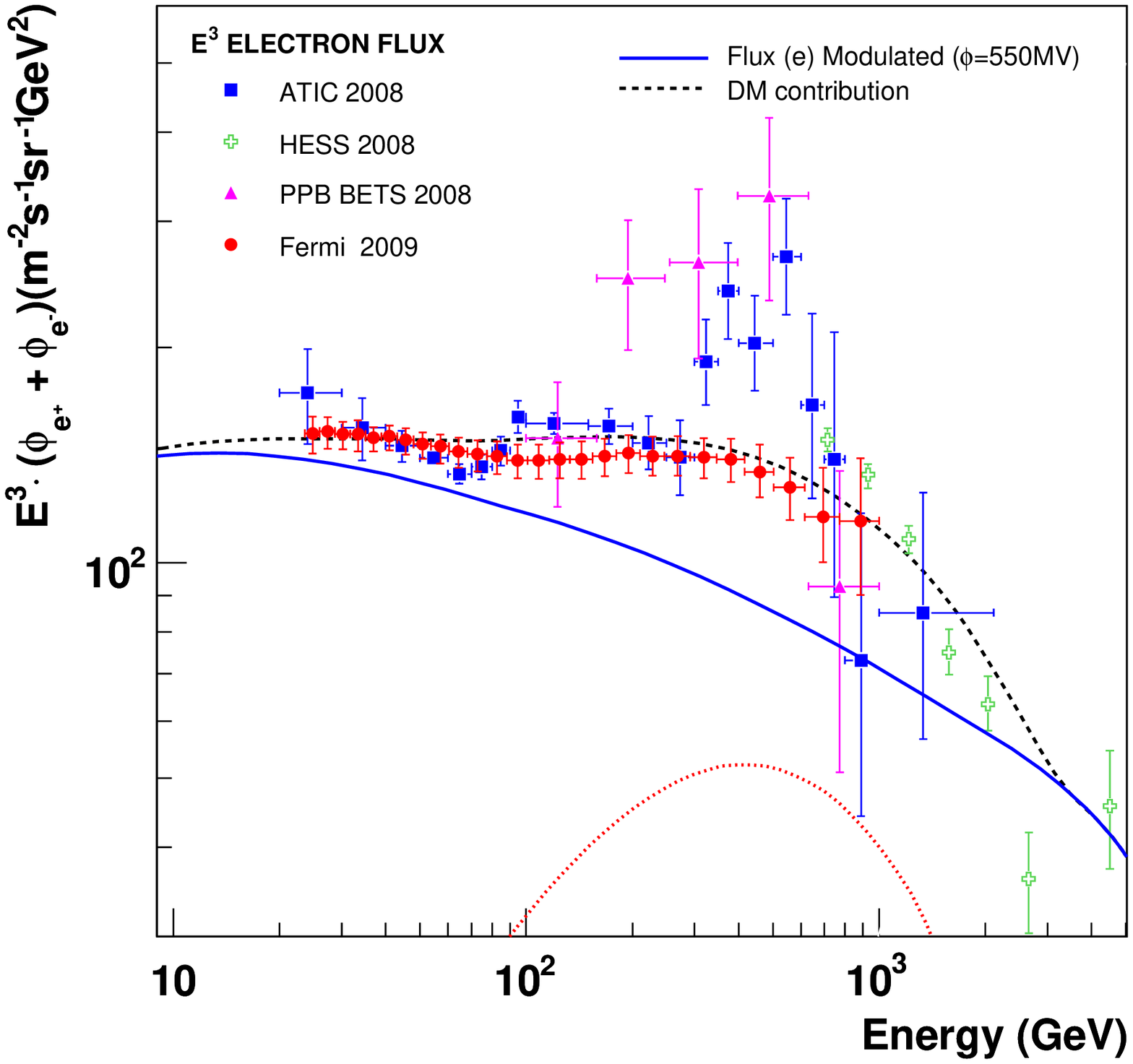}}
   \subfigure[]{\includegraphics[width=0.47\textwidth]{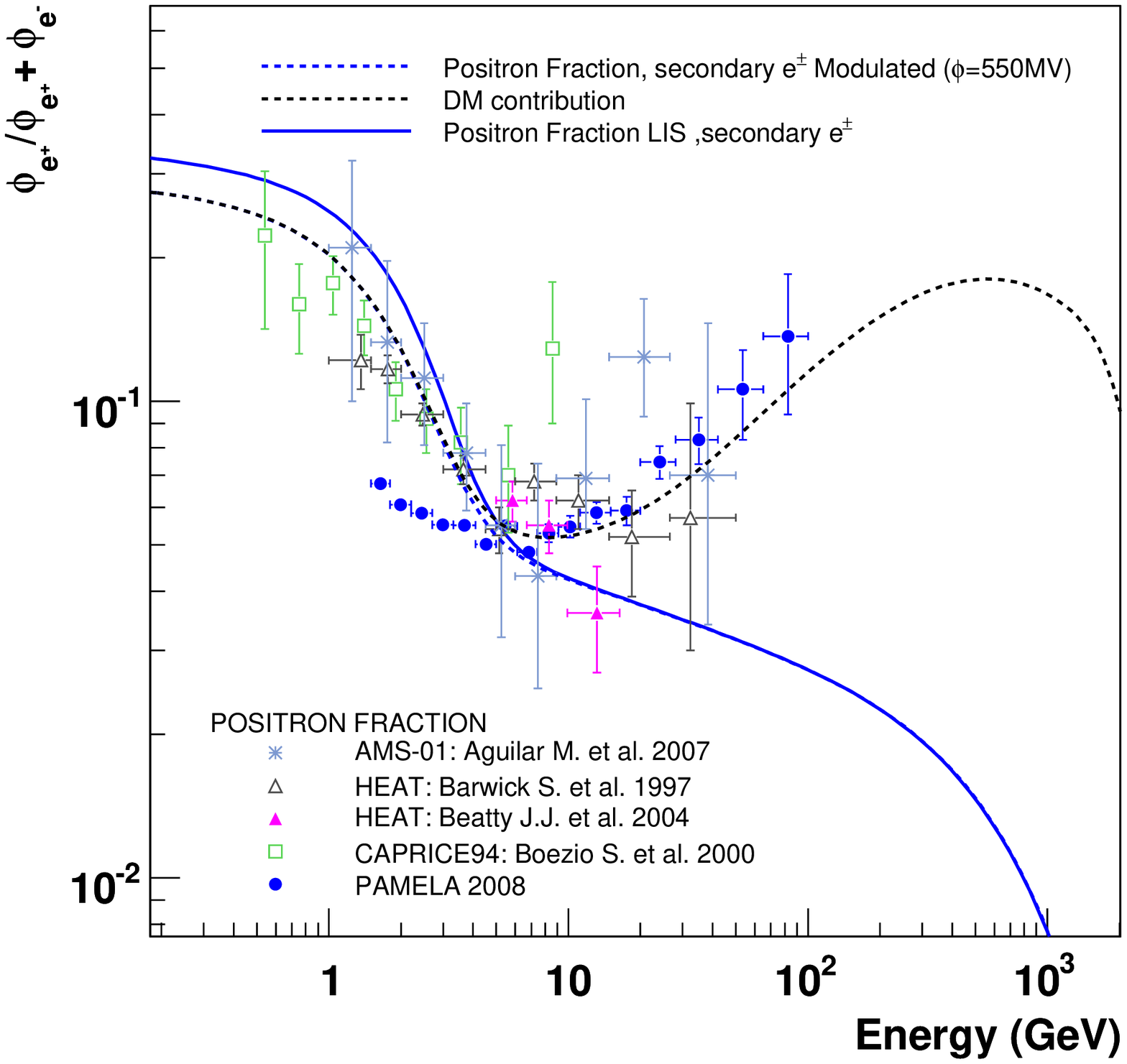}}
 \label{fig:fig1}
 \caption{ DM clumpy halo contribution to the electron spectrum as described in the text. Annihilation into $\tau^{+}\tau^{-}$ with a cross section 
 $\langle\sigma v\rangle=4.65\rm{x}10^{-24}\rm{cm^{3}/s}$. The considered DM mass  $M_{DM}=3.6$ TeV and concentration parameter c=1.6. MED diffusion setup.
     (a): $E^{3}\cdot (e^{+}+e^{-})$ (b): Positron Fraction }
  \end{figure}

As previously stated, PAMELA data favours leptophilic DM, ie , candidates which annihilation products are predominantely leptons. The case of direct annihilation into electron and
positron pairs can provide good fits to the ATIC data but it is excluded if we take into account the Fermi results due to a sharp drop at the end-point (see e.g. 
\cite{Meade et al (2009)}). Therefore we are left
with annihilations into $\mu^{\pm}$ and $\tau^{\pm}$ which provide a much softer injection spectrum and can accomodate both PAMELA and Fermi. For such injection spectrums, large
annihilation rates are required.
The astrophysical boost factor, i.e, boost based on the presence of clumpiness, has been proved to be inssufficient to account for the required normalization to adjust the data. In
this respect, proposals of particle physics boost factors as velocity dependent cross sections, have been recently considered.\\
In this work our purpose is to evaluate the expected
anisotropy resulting from sources that reproduces the observed abundances,so we will assume the required
normalization to fit the data as a result of any of the considered mechanisms that can boost the annihilation rate. Here, we consider a DM candidate that
annilates into $\tau^{\pm}$ with a mass of 3.6 TeV (Fig. \ref{fig:fig1}).

We also address the posibility of a nearby DM clump being the responsible of the bulk of positrons found in the Fermi,ATIC and PAMELA data.
For this purpose, we treat here the clump as a point-like object at a given distance \emph{d} contributing to the source term with luminosity \emph{L} in the same
fashion as already done by \cite{Brun et al. (2009)}. This possibility has been found to be very unlikely unless the Sommerfeld effect is at play, but such a source would imprint
a signature in the electron arrival direction that could constitute a signal of dark matter annihilation given the absence of pulsars in the neighbourhood. In this sense, the evaluation of such a signal makes the study
meaningful. The expected flux for a clump of DM annihilating into $\tau^{\pm}$ or $\mu^{\pm}$ with masses around 3 TeV and 2 TeV respectively can reproduce the Fermi data with the
observed drop by Hess , and with
slightly different normalizations the PAMELA data. As a posibility, we show the flux produced by the annilation in the $\tau^{\pm}$ channel of a DM clump located at 0.9 Kpc with a
DM mass of 3.6 TeV . In this case , as in a pulsar scenario, a MED diffusion setup is apropiate , as the MIN and MAX cases show prominent bumps in the spectrum at high and low energies respectively, that
are not observed in the Fermi data.

 \begin{figure}[!ht]
 \centering
   \subfigure[]{\includegraphics[width=0.47\textwidth]{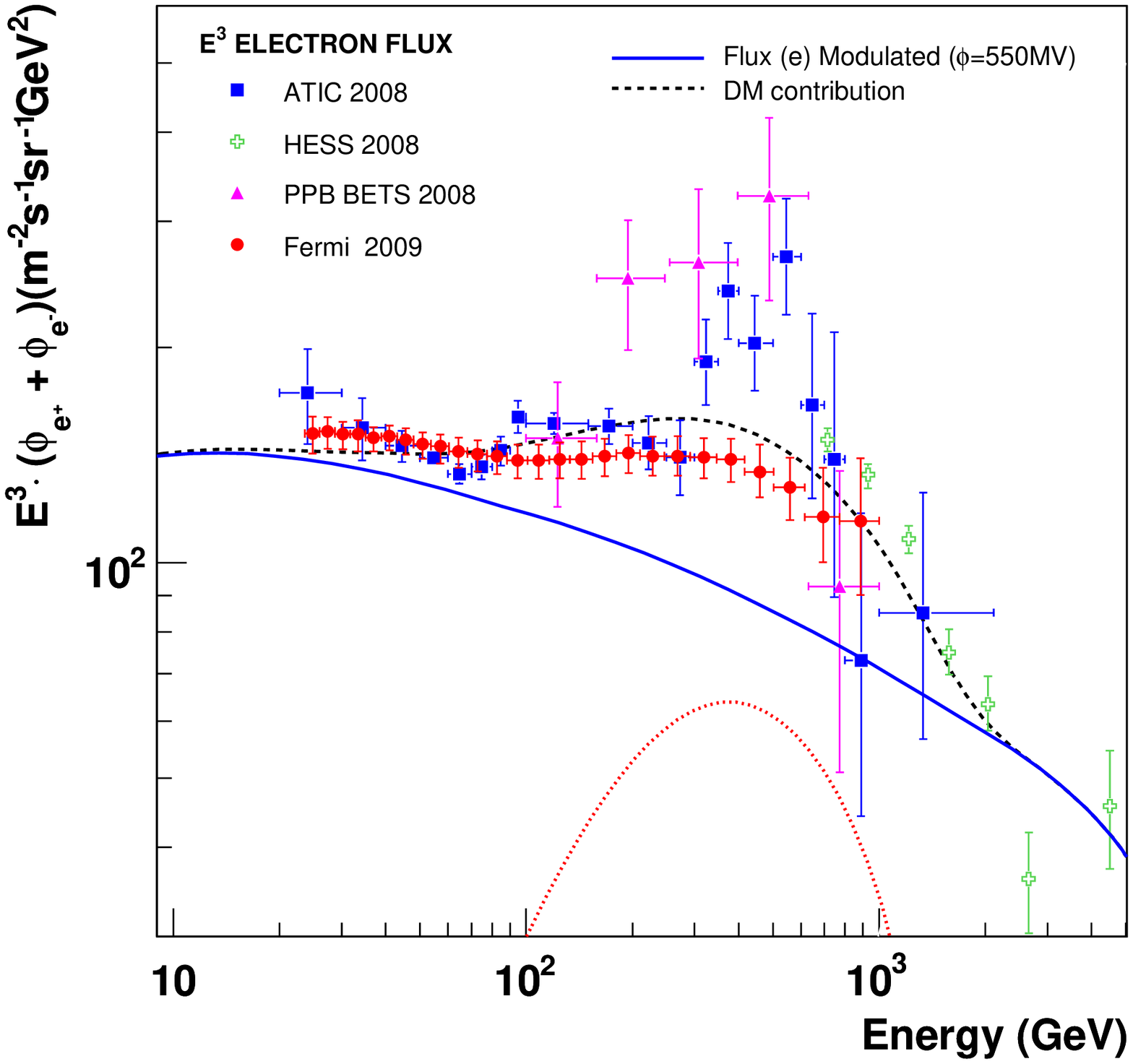}}
   \subfigure[]{\includegraphics[width=0.47\textwidth]{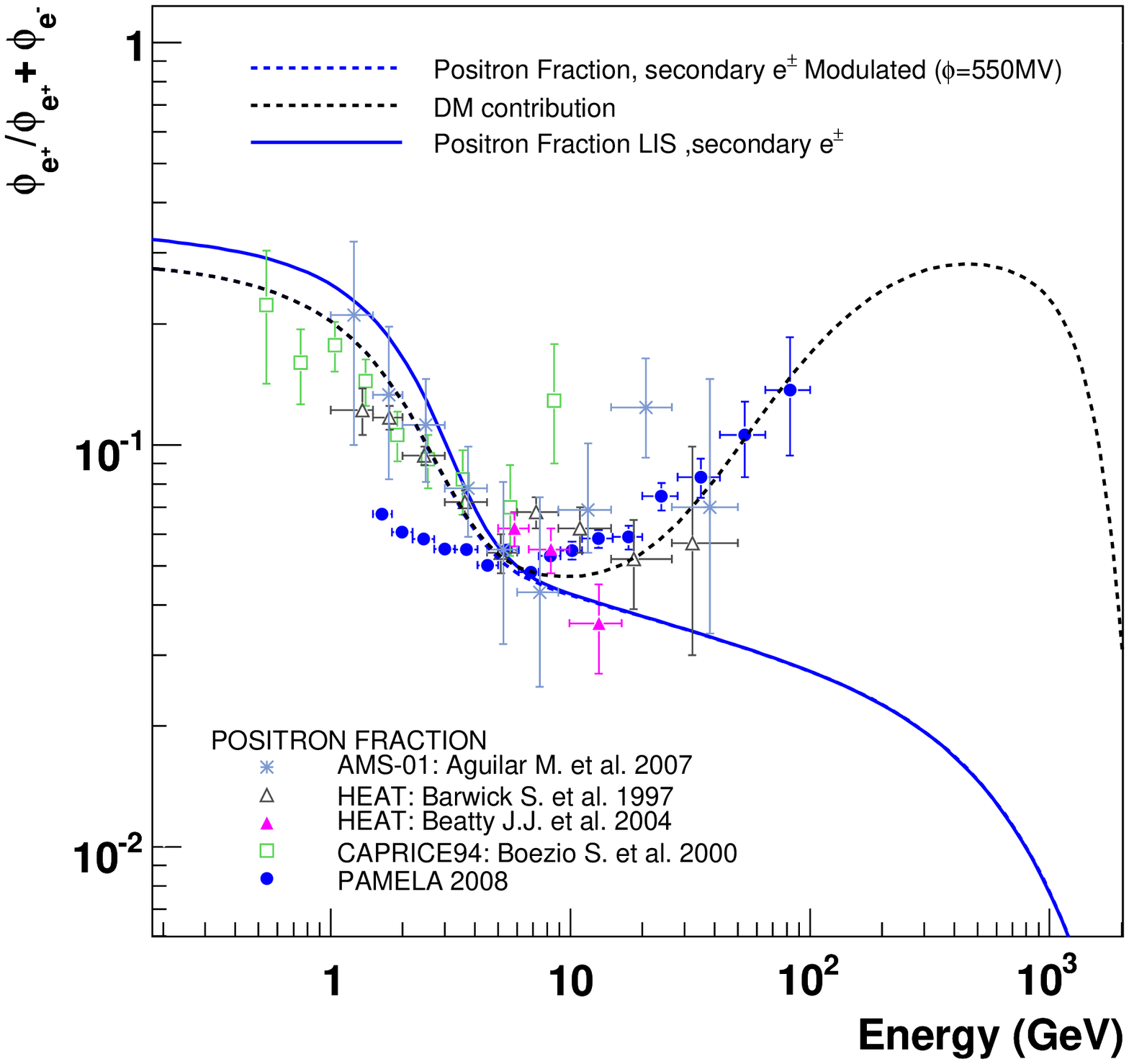}}
 \label{fig:fig2}
 \caption{ DM clump situated at d=0.9Kpc. MED diffusion setup and luminosity $L=1.0\rm{x}10^{9} M_{\odot}^{2}pc^{-3}$. We assume a DM candidate of mass 3.6 TeV that annilates into $\tau^{+}\tau^{-}$
    with cross section $\langle\sigma v \rangle=3\rm{x}10^{-25}\rm{cm^{3}/s}$ (a): $E^{3}\cdot (e^{+}+e^{-})$ (b): Positron Fraction }
  \end{figure}

\section{Anisotropies}

 Large scale anisotropies of CR have been measured so far to be less than 1\% \cite{Amenomori et al. (2006)}, but it is known that if it were produced by sources with some spatial structure, small anisotropies should be
 present in the arrival directions and could be correlated to the potential sources, whether it be known or not. In the case of high energy electrons, very light charged particles, diffusion
 competes with large energy losses resulting in relatively short paths $\bf${O}(Kpc), so it is expected that we can use them to sample only sources within a certain distance and
 age.
 The cosmic ray intensity can, in general, be expanded over the celestial sphere in spherical harmonics. At first order, when we have a clear directionality, we have a dipole
 anisotropy as could be the case of a single source dominating the spectrum. In this case, the intensity can be expressed as
   
\begin{equation}
  I(\theta)=\bar{I} + \delta\bar{I}\cos\theta
\end{equation}

 where $\bar{I}=1/2(I_{max}+I_{min})$ being this maximum and minimum intensities related to a forward-backward measurement.
 
 The calculation of the anisotropy produced by a single source has been carried out by \cite{Mao et al. (1971)} and is given by: 
  
\begin{equation}
  \delta_{i}=\frac{3D}{c}\frac{\mid\nabla N\mid}{N}
\end{equation}

Where D(E) is de diffusion coefficient and N is the electron number density.

In order to detect a statistically significant anisotropy , at the $2\sigma$ level, $\delta> 2\sqrt{2}(N_{evts})^{-1/2}$ where $N_{evts}$ is the number of events above a given
energy threshold ,that is a function of the detector
acceptance and the exposition time. $N_{evts}(E>E_{th})=\int_{E_{th}}\phi(r,E)\cdot Acc\cdot T_{exp}dE$.

Estimates of the expected anisotropy in the case of a dominant pulsar have been previously shown in e.g. \cite{Busching et al. (2008)},\cite{Hooper-Blasi-Serpico (2008)} but they fail to take into account the possible effects that could arise
from a realistic collection of pulsars, there could  be e.g. systematic cancellations.
We have calculated the anisotropy in the case of a collection of pulsars taken from the ATNF catalogue and , for the first time, in the framework of dark matter annihilation of clumps thoughout the
halo,both in the case of a dominating point-like source or a distribution of clumps.
In the presence of an isotropic background plus a number of contributing sources to the total flux, the expected dipole anisotropy is given by :

\begin{equation}
  \delta=\frac{\sum \phi_{i}(E,r,t)\langle\delta_{i}\hat{r}_{i}\hat{n}_{i}\rangle}{\phi_{T}}
\end{equation}

where the index i runs over all the discrete sources that contribute to the full dipole. The product $\langle\delta_{i}\hat{r}_{i}\hat{n}_{i}\rangle$  represents the projection of the individual dipole over
the direction of maximum intensity that is energy dependent and $\phi_{T}(E,r,t)$ is the total flux observed at Earth.
The projection can be easily calculated taking into account the angular separation $\theta$ of two sources on the celestial sphere given by:

\begin{eqnarray}
  \theta=\arctan
  \left(\frac{\sqrt{A1 + A2^{2}}}
  {\sin\delta_{1}\sin\delta_{2}+\cos\delta_{1}\cos\delta_{1}\cos(\alpha_{2}-\alpha_{1})}\right)
\end{eqnarray}

\begin{figure}[!hb]
\begin{center}
\includegraphics[width=0.7\columnwidth]{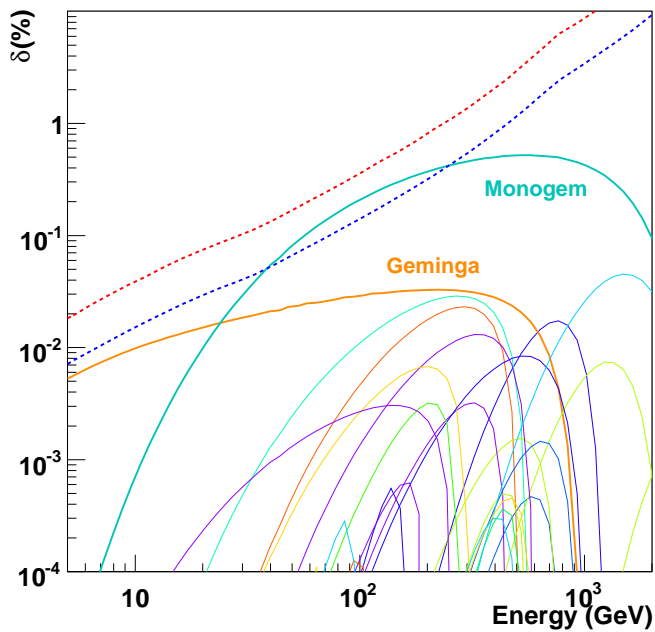}
\caption{Individual dipole anisotropy in the electron + positron spectrum for every pulsar considered in Fig. \ref{pulsars} (a) for a MED diffusion setup.Also shown the Fermi sensitivity to such an anisotropy at the 2 and 5
$\sigma$ CL in 5 years.}
\label{fig:indivani}
\end{center}
\end{figure}

\begin{figure}[!hb]
\begin{center}
\includegraphics[width=0.7\columnwidth]{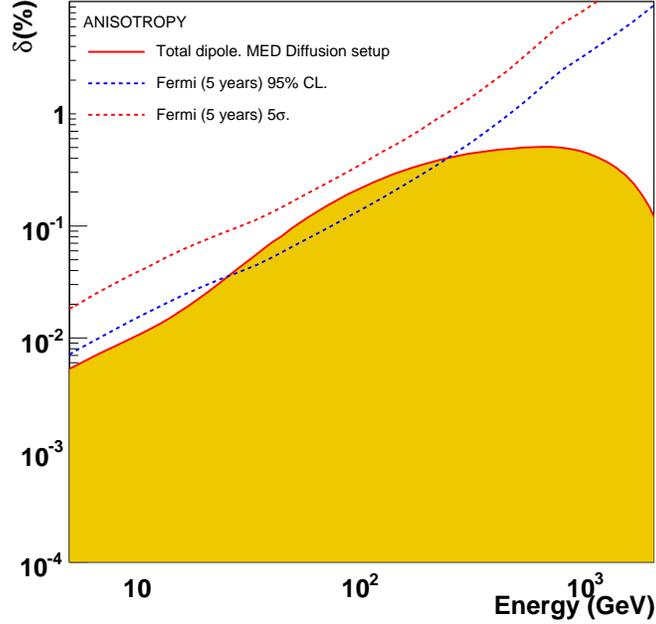}
\caption{Full dipole anisotropy in the electron + positron spectrum from collection of pulsars as considered in Fig. \ref{pulsars} for a MED diffusion setup.Also shown the Fermi sensitivity to such an anisotropy at the 2 and 5
$\sigma$ CL in 5 years.}
\label{fig:fig3}
\end{center}
\end{figure}

Where $A1=\cos^{2}\delta_{2}\sin^{2}(\alpha_{2}-\alpha_{1})$ , $A2=\cos \delta_{1}\sin\delta_{2}-\sin\delta_{1}\cos\delta_{2}\cos(\alpha_{2}-\alpha_{1})$ and the right ascension
and declination are denoted by $\alpha_{i}$  and  $\delta_{i}$.\\
In this work, we evaluate the anisotropy (Fig. \ref{fig:indivani}) produced by the collection of pulsars considered in Fig. \ref{pulsars} (a) for a MED diffusion setup.\\ 
The dipole anisotropy will change direction depending on the energy, but the main contributor to the full dipole is shown to be Monogem above a few tens of GeV up to the TeV and
in second place Geminga.
A big contribution  to the anisotropy (and the positron flux) can be expected also at higher energies by younger objects like Vela or CTA1, although it is still unclear if such an object
can produce a sizeable amount of electrons due to the PWN confinement.
The full dipole (Fig. \ref{fig:fig3}) will be given by the projection of the individual anisotropies in the direction of the maximum intensity at each
energy, resulting in a clear signal in the direction of Monogem at energies above 20 GeV. The contribution due to Geminga has an addition effect to the full
anisotropy due to the relative position between it and Monogem, so, changes in their individual contributions to the flux , codified for instance in the pair conversion efficiency, should not change the anisotropic pattern.
Contributions from other pulsars don't result in a systematic addition of their signal to the full anisotropy into a particular direction, due to their spatial
distribution, so they constitute a kind of isotropic background.
In this scenario, measurements of a possible privileged incomimg direction should point out an excess in the Monogem/Geminga direction, roughly oposite to the direction of the MW.
It is possible however that we are observing the contribution of some yet undiscovered pulsars that could show up in a potential study of anisotropies. In this respect, searches
for gamma ray sources as the one conducted by Fermi will help to support or disfavour the test.

In the case of a number of clumps contributing to the full dipole, as in the case of Diemands simulation, due to the symmetry of the clump distribution we would expect a dipole
anisotropy in the direction of the Milky Way (MW) center. The expected dipole anisotropy will be given by 

\begin{equation}
  \delta=\frac{1}{\phi_{T}}\int \phi(E,\bf{r})\langle\delta_{clump}(\bf{r})n\rangle d\bf{r}^{3}
\end{equation}

where $\langle\delta_{clump}(r)n\rangle=\frac{6D}{cD_{diff}^{2}(E)}(r_{\odot}-r\cos\varphi)$ is the projection of the clump contribution to the full dipole in the direction of the
MW center and r,$\varphi$ are the cylindrical coordinates of the clump distribution.

If we introduce the clump distribution we have used in the previous section, the expected dipole anisotropy for the direct annihilation channel will be :

%\begin{eqnarray}
%  \delta(E)=\frac{1}{\phi_{T}}\frac{\beta c}{8\pi}\frac{\langle\sigma v\rangle}{m^{2}_{DM}}\int^{M_{max}}_{M_{min}}d\log(M/M_{\odot})\frac{dn_{D}}{d\log(M/M_{\odot})}&& \nonumber   \\%
%  f^{2}_{NFW}(c,M)\int rdrd\phi dz\frac{6D(E)}{cD_{diff}^{2}(E)}(r_{\odot}-r\cos\phi)G(E,r)&&
%\end{eqnarray}

\begin{eqnarray}
  \delta(E)=\frac{1}{\phi_{T}}\frac{\beta c}{8\pi}\frac{\langle\sigma v\rangle}{m^{2}_{DM}}\int rdrd\varphi dz\frac{6D(E)}{cD_{diff}^{2}(E)}(r_{\odot}-r\cos\varphi)G(E,r)&& \nonumber   \\%
  \int^{M_{max}}_{M_{min}}d\log(M/M_{\odot})\frac{dn(r,M)}{d\log(M/M_{\odot})} f^{2}_{NFW}(c,M)&&
\end{eqnarray}

\begin{figure}[!h]
\begin{center}
\includegraphics[width=0.7\columnwidth]{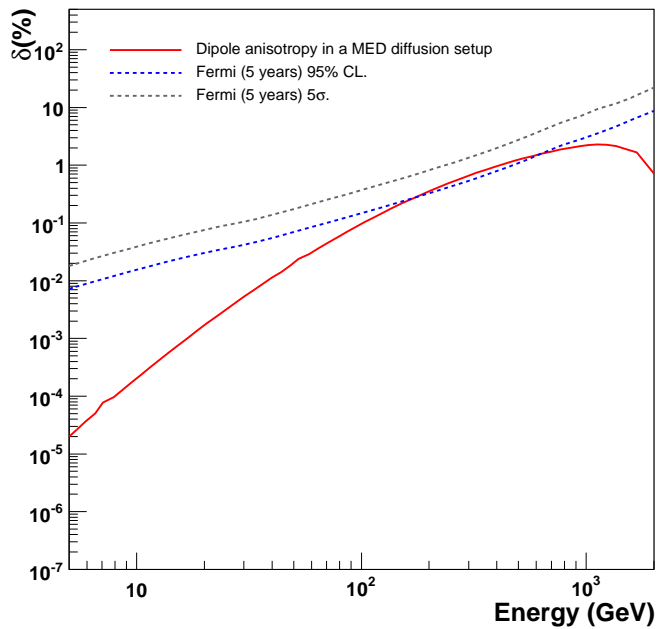}
\caption{Dipole anisotropy in the electron + positron spectrum from the DM point-like source as considered in Fig. \ref{fig:fig2}.Also shown the Fermi sensitivity to such an anisotropy at the 2 and 5
$\sigma$ CL in 5 years.}
\label{fig:fig4}
\end{center}
\end{figure}

\begin{figure}[!h]
\begin{center}
\includegraphics[width=0.7\columnwidth]{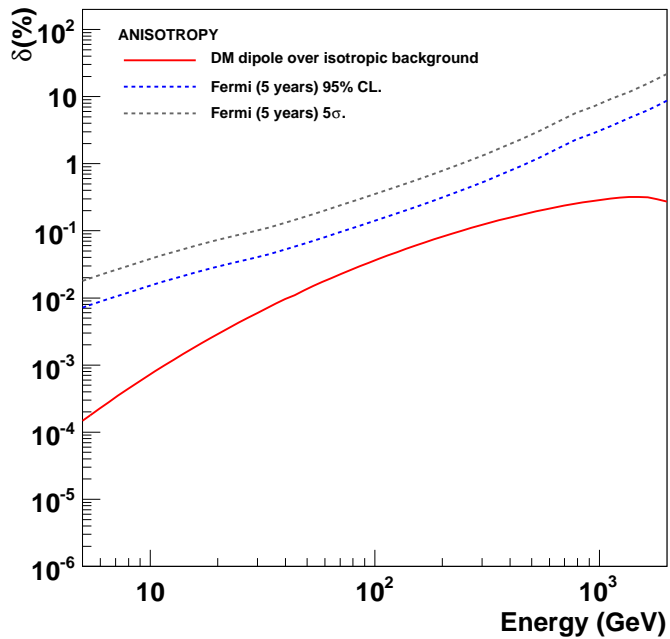}
\caption{Dipole anisotropy in the electron + positron spectrum from a distributed DM source considered in Fig. \ref{fig:fig1}.Also shown the Fermi sensitivity to such an anisotropy at the 2 and 5
$\sigma$ CL in 5 years.}
\label{fig:fig5}
\end{center}
\end{figure}

It must be noted that in case we have an initial source distribution f(E) of electrons as the one that comes from annihilation through the $\tau^{+}\tau^{-}$ channel, an equivalent
result is obtained
after the convolution of the source term  with the $\nabla N$ term of the anisotropy definition.  The resulting signal (Fig. \ref{fig:fig5}) for the parametrization considered in Fig. \ref{fig:fig1} , would not reach the Fermi sensitivity for a 5 years survey as the 
Fermi spectrum is too flat. On the contrary, fits to the much bumpier ATIC data would provide a signature at the 2$\sigma$ C.L. towards the MW center providing a hint to the origin of the positron
excess.\\
We can also consider a DM point source as the responsible of the bulk of PAMELA electrons as could be the case of a large DM clump \cite{Brun et al. (2009)} or minispikes around an intermediate mass black hole
\cite{Bringmann et al. (2009)}. In this case
the expected anisotropy is reduced to $\delta =\frac{\phi(E,r)\delta_{point source}}{\phi_{T}}$ where $\delta_{point source}=\frac{6D(E)d}{cD_{diff}^{2}(E)}$ being d the distance to
the source (Fig. \ref{fig:fig4}). \\
Although the probability of finding such a bright clump in our neighbourhood is rather small, there are some scenarios, with the Sommerfeld effect at play, where this probability
can be boosted up to a 15\%. In this case, the anisotropy would exceed the $2\sigma$ level pointing toward the existence of a dominant source. This measurement should also be
complemented with searches for gamma ray emission to achieve a consistent prediction.

\section{Conclusions}
%\label{Conclu}

 A number of posibilities have been proposed as potential candidates to account for the positron excess . From the standard astrophysical point of view, pulsars seem to be the
 most promising candidates. Just considering the already known
 gamma-ray pulsars is enough to explain the spectrum for reasonable model assumptions, but the features in the ATIC data require quite a large conversion factor into pairs for
 the considered model . On the other hand, Fermi data shows a much flatter spectrum that can be reproduced with a reasonable set of pulsar parameters.
  If we consider dark matter annihilation, the
 contribution from the clumpy halo can reproduce the observed patterns once we have the required normalization issue solved by means of particle physics boosts on the thermal
 averaged cross section.
 One way or the other, the measured spectrum can be reproduced and no clear signatures can be found to distinguish between a dark matter scenario or pulsars, as the spectral shape
 seems to lack information of the origin of the electrons. In order to distinguish between the proposed candidates,
 the study of anisotropies is proposed , as a not performed yet test, that can provide useful information. In this work we have analysed the expected anisotropy from a
 configuration of pulsars as in the ATNF catalogue. We have also derived the anisotropy in a general dark matter scenario, both in case we have a very bright point source and in the
 case of a clumpy distribution as ilustrated by N-body simulations . For the DM point-like source and the pulsar scenarios, the Fermi observatory should be able to detect a dipole
 anisotropy at 2 sigma CL in five years at least . On the other hand, we would expect an excess towards the MW center in the case of a clumpy halo as the main contributor to the
 primary positron flux, but fits to the Fermi
 spectrum doesn't provide enough events in a 5 years survey. In contrast, the case of the ATIC spectrum would open the posibility of direct annihilation into $e^{+}e^{-}$ implying
 a much harder spectrum. Even annihilations through $\tau^{\pm},\mu^{\pm}$ channels would imprint a signature in the electron anisotropy detectable by Fermi that are not expected to be observed
 in view of the the Fermi data.\\
 Anisotropies have been shown to contribute in a valuable way to disentangle the positron excess problem, nonetheless there are still a lot of theoretical uncertainties (e.g. the
 dark matter halo distribution or the mechanism of pair production in pulsars). It must be borne in mind that we have not taken into account the proper motion of the pulsars or
 even that of the Dark Matter clumps, that could result in an enhancement or suppression of the anisotropy. Moreover, the case of a dipole anisotropy toward a pulsar cannot exclude a dark matter scenario,
 as it is possible to have a large Dark Matter cloud in the same direction masking the signal. In any case, the precise study of electron anisotropies should be conducted together with
 pulsar surveys to help to discriminate between astrophysical and more exotic sources.

\section{Acknowledgements }
I would like to acknowledge the support from Javier Berdugo, Carlos Ma\~n\'a, Jorge Casaus and Carlos Delgado for their useful suggestions and endless patience and Aurelio Carnero for the
proofreading. I would also like to thank D.Grasso, D.Gaggero and G.DiBernado for their useful remarks and the anonymous referee for his helpful comments.
This work is partially supported by the Centro de Investigaciones Energeticas, Medioambientales y Tecnologicas (CIEMAT) and the MICINN FPI Grant number BES-2007-151967.


\begin{thebibliography}{00}


\bibitem{Boezio et al. (2008)} Observation of an anomalous positron abundance in the cosmic radiation Boezio M. on behalf of Pamela's collaboration Nature, arXiv:0810.4995v1 astro-ph
\bibitem{ATIC}J. Chang et al., Nature 456, 362 (2008).
\bibitem{PPBBETS} S. Torii et al., submitted to Astropart. Phys arXiv:0809.0760 [astro-ph].
\bibitem{FermiData} Measurement of the Cosmic Ray e+ plus e- spectrum from 20 GeV to 1 TeV with the Fermi Large Area Telescope, FERMI/LAT collaboration, arXiv:0905.0025.
\bibitem{Heat} S. W. Barwick et al. [HEAT Collaboration], Astrophys. J. 482, L191 (1997).[arXiv:astro-ph/9703192]; J. J. Beatty et al., Phys. Rev. Lett. 93, 241102 (2004)[arXiv:astro-ph/0412230].
\bibitem{AMS} M. Aguilar et al. [AMS-01 Collaboration], Phys. Lett. B 646, 145 (2007) [arXiv:astro-ph/0703154].
\bibitem{Delahaye et al. (2008)} Galactic secondary positron flux at the Earth T. Delahaye, F. Donato, N. Fornengo, J. Lavalle, R. Lineros, P. Salati, and R. Taillet,[arXiv:0809.5268].
\bibitem{Cirelli et al. (2008)} Model-independent implications of the e+, e-, anti-proton cosmic ray spectra on properties of Dark Matter Marco Cirelli, Mario Kadastik, Martti Raidal, Alessandro Strumia [arXiv:0809.2409v2].
\bibitem{solut} The origin of cosmic rays.V.L. Ginzburg and S.I. Syrovatskii,  Pergaemon,Oxford,1964.
\bibitem{Strong et al. (2007)} Cosmic-ray propagation and interactions in the Galaxy Andrew W. Strong, Igor V. Moskalenko, Vladimir S. Ptuskin,  [astro-ph/0701517].
\bibitem{Green}Positrons from particle dark-matter annihilation in the Galactic halo: propagation Green's functions. I. V. Moskalenko, A. W. Strong , Phys.Rev. D60 (1999) 063003.
\bibitem{Edsjo}Positron propagation and fluxes from neutralino annihilation in the halo. Baltz, E. A. \& Edsjo, J. 1999, Phys. Rev. D, 59, 023511
\bibitem{Busching et al. (2008)} A cosmic-ray positron anisotropy due to two middle aged nearby pulsars, I.Busching, O.C. de Jager, M.S. Potgieter and C. Venter, The AstrophysicalJournal,678:L39-L42,2008 May.
\bibitem{Hooper-Blasi-Serpico (2008)} Pulsars as the Sources of High Energy Cosmic Ray Positrons, Dan Hooper,Pasquale Blasi and Pasquale Dario Serpico, FERMILAB-PUB-08-429-A.
\bibitem{Yuksel et al. (2008)}TeV Gamma Rays from Geminga and the Origin of the GeV Positron Excess, [arXiv:0810.2784v3].
\bibitem{Harding et al.}A. K. Harding and R. Ramaty, in International Cosmic Ray Conference (1987), vol. 2 of International Cosmic Ray Conference, pp. 92.
\bibitem{Chi et al} X. Chi, K. S. Cheng, and E. C. M. Young, ApJ 459, L83+ (1996).M
\bibitem{Gao et al. (2008)} Possible Contribution of Mature gamma ray Pulsars to Cosmic ray Positrons. Chin.J.Astron. Astrophys. Vol. 8 (2008),No 1, 87-95.
\bibitem{Profumo (2008)} Dissecting Pamela (and ATIC) with Occam's Razon: Existing,well known Pulsars naturally account for the "anomalous" Cosmic-Ray Electron and Positron Data
Profumo Stefano [arXiv:0812.4457].
\bibitem{Grasso et al. (2009)} On Posible Intepretations of the high Energy Electron-Positron Spectrum measured by the Fermi Large Area Telescope. D. Grasso, S. Profumo, A.W. Strong, L. Baldini, R. Bellazzini, E.D. Bloom, J. Bregeon, G. Di Bernardo, D. Gaggero, N. Giglietto, T. Kamae, L. Latronico, F. Longo, M.N. Mazziotta, A.A. Moiseev, A. Morselli, J.F. Ormes, M. Pesce-Rollins, M. Pohl, M. Razzano, C. Sgro, G. Spandre, T.E. Stephens
[arXiv:0905.0636]
\bibitem{Zhang et al. (2001)} L.Zhang and K.S.Cheng,A\&A 368,1063 (2001).
\bibitem{Manchester et al. (2004)} The ATNF Pulsar Catalogue, Manchester, R. N., Hobbs, G. B., Teoh, A. and Hobbs, M., Astron. J., 129, 1993-2006 (2005) [astro-ph/0412641].
\bibitem{Zhang et al. (1997)} L. Zhang and K.S. Cheng, Astrophys.J. 487,370 (1997).
\bibitem{Kawanaka et al. (2009)} Cosmic-Ray Electron Excess from Pulsars is spiky or smooth? Continuous and multiple electron/positron injections.Norita Kawanaka, Kunihito Ioka and Mihoko M. Nojiri.[arXiv:astro-ph/0903.3782v1].
\bibitem{Atoyan et al. (1995)} Electrons and positrons in the galactic cosmic rays, A. M. Atoyan, F. A. Aharonian and H. J. Volk, Phys.Rev. D 52, 3265, (1995).
\bibitem{D.Malyshev et al (2009)} Pulsars versus Dark Matter Interpretation of ATIC/PAMELA , D.Malyshev,I. Cholis and J. Gelfand 2009.[arXiv:0903.1310].
\bibitem{Diemand et al. (2006)} Juerg Diemand, Ben Moore and Joachim Stadel, 2005, Nature 433,389.
\bibitem{Cumberbatch et al. (2006)} Local dark matter clumps and the positron excess, Daniel Cumberbatch and Joseph Silk, Mon.Not.Roy.Astron.Soc.374:455-465,2007 [arXiv:astro-ph/0602320v3]
\bibitem{Meade et al (2009)} Dark Matter Interpretations of the Electron/Positron Excesses after FERMI. Patrick Meade, Michele Papucci, Alessandro Strumia, Tomer Volansky . IFUP-TH-2009-9, May 2009. [arXiv:0905.0480v1 [hep-ph]]
\bibitem{Brun et al. (2009)} The cosmic ray lepton puzzle in the light of cosmological N-body simulations , Pierre Brun, Timur Delahaye,Jurg Diemand,Stefano Profumo and Pierre Salati [arXiv:0904.0812v2]
\bibitem{Amenomori et al. (2006)} Anisotropy and Corotation of Galactic Cosmic Rays, Amenomori et al. Science (2006) Vol. 314. no. 5798, pp. 439 - 443.
\bibitem{Mao et al. (1971)} Anisotropy and Diffusion of Cosmic Ray Electrons , Mao, C.Y.,  and Shen ,C.S Chinese Journal of Physics, Vol 10, No1, April 1972.
\bibitem{Hooper et al. (2008)} The PAMELA and ATIC Excesses From a Nearby Clump of Neutralino Dark Matter ,Dan Hooper, Albert Stebbins and Kathryn M.Zurek., [arXiv:hep-ph/0812.3202].
\bibitem{Bringmann et al. (2009)} Intermediate Mass Black Holes and Nearby Dark Matter Point Sources: A Myth-Buster ,Torsten Bringmann,Lulien Lavalle and Pierre Salati, [arXiv:astro-ph/0902.3665v1].
\bibitem{Fermi} Measuring 10-1000 GeV Cosmic Ray Electrons with GLAST/LAT, Alexander A.Moiseev, Jonathan F.Ormes, Igor V.Moskalenko,Proc. 30th ICRC (Merida), 2, 449-452 (2007),[arXiv:0706.0882v1]
\end{thebibliography}
\end{document}